\documentclass[11pt]{article}

\usepackage[final]{acl}

\usepackage{times}
\usepackage{latexsym}

\usepackage[T1]{fontenc}

\usepackage[utf8]{inputenc}

\usepackage{microtype}

\usepackage{inconsolata}

\usepackage{graphicx}

\usepackage{amsmath}
\usepackage{booktabs}

\title{Who Plays Which Role When? Communication Role Dynamics for Peer Recognition and Team Performance Prediction}

\author{
Yifan Song\textsuperscript{1},
Wenxuan Wendy Shi\textsuperscript{2},
Brian P. Bailey\textsuperscript{1},
Tal August\textsuperscript{1}
\\
\textsuperscript{1}University of Illinois Urbana-Champaign 
\\
\textsuperscript{2}California Polytechnic State University, Pomona
\\
\texttt{\{yifan33, bpbailey, taugust\}@illinois.edu},\quad\texttt{wendyshi@cpp.edu}
}

\begin{document}
\maketitle

\begin{abstract}
Team roles offer an interpretable lens on collaboration, yet computational studies of roles often rely on domain-specific personas or data-driven clustering rather than theory-grounded taxonomies. We operationalize a taxonomy of eight communication roles grounded in education literature and annotate a corpus of 6,307 Slack messages from 55 students across 18 teams in a semester-long computer science course project. We evaluate whether LLMs can approximate expert labels, enabling scalable, taxonomy-driven role annotation. Using these role labels, we characterize role dynamics over teams' lifecycles, finding that different roles peak at different moments and that students enact a more diverse set of roles as projects progress. To evaluate the utility of our role constructs, we use them to predict peer recognition, outperforming lexical, conversational, and LLM-prompting baselines. To assess generalizability beyond the educational context, we apply the same role constructs to a public dataset (DeliData) to predict team performance improvement after deliberation, again exceeding prior performance.
\end{abstract}

\section{Introduction}

Effective teamwork depends not only on what a team produces, but also on how members communicate, such as introducing ideas, coordinating work, and managing disagreement. A useful lens for describing these interaction patterns is \textit{team roles}. Behavioral and organizational science conceptualizes roles as complementary communicative functions enacted in interaction (e.g., an \textit{initiator} who proposes new ideas, or an \textit{arbitrator} who solves disagreements), rather than as fixed personality traits \citep{benne_functional_1948, meredith_belbin_management_2011}. This perspective naturally raises a core question for computational analysis of team communication: who enacts which roles, when, and how do these patterns relate to collaboration quality?

Prior work has modeled functional roles for meeting participants from simple speech features \citep{banerjee-rudnicky-2006-say}, or used behavioral patterns (e.g., turn-taking) to predict "latent" roles and team outcomes \citep{yang-etal-2015-weakly}. Recently, large language models (LLMs) have renewed interest in roles through human-agent and agent-agent collaboration, where agents are assigned explicit social roles in games or simulated organizations \citep{lan-etal-2024-llm, li-etal-2025-advancing}. However, roles are often operationalized as domain-specific personas or functions (e.g., \textit{developer} vs.\ \textit{manager}), rather than grounded to theory-driven communicative functions. In addition, most prior work studies roles in large-scale (e.g., Wikipedia), controlled (e.g., crowdsourced), or synthetic (e.g., agent simulation) contexts \citep{maki-etal-2017-roles, litman-etal-2016-teams, lu-etal-2024-large}. While these contexts offer scale and control, they differ from the close-knit, long-lived teams common in educational, research, and organizational settings \citep{work_teams_2017}, and often miss how real teams evolve over time.

In this paper, we ground team roles in an authentic longitudinal setting: an in-person, semester-long computer science course project in which teams relied on Slack for day-to-day coordination. We collect a dataset of 6,307 messages from 55 students across 18 teams over eight project deliverable windows. We operationalize a role taxonomy grounded in the education literature \citep{Nestsiarovich2020TeamRA} and develop an expert annotation protocol for labeling the eight roles based on the taxonomy. We then evaluate an LLM-as-annotator setup as a scalable approximation to expert annotation, with the best model yielding good overall agreement with expert labels.

Using these role labels, we provide descriptive analyses of role dynamics---the longitudinal evolution of an individual's communicative functions---including role prevalence and trajectories over the project lifecycle. For example, work-related roles like \textit{explorer} and \textit{facilitator} peak during the intensive implementation phase, while \textit{gatekeeper} increases substantially near the end of the project during the busiest and most stressful period. We also observe a progressive increase in the number of unique roles an individual enacts simultaneously as project complexity grows.

To evaluate the usefulness of these role constructs, we first predict peer recognition, an individual-level performance measure derived by normalizing peer-evaluation scores. We show that role-based features improve predictive performance over lexical, conversational, and zero-shot LLM baselines, yielding about 5--10\% improvement. To further validate generalizability beyond the educational context, we use the same LLM-annotated role constructs to predict team performance gain in a public dataset \citep[DeliData,][]{karadzhov2023delidata} of group deliberation dialogues. Combining our role features with conversational statistics outperforms all baselines reported in the original work.

Our contributions include:
\begin{itemize}\setlength{\itemsep}{1pt}
    \item We show that a theory-grounded role taxonomy can be used to reliably annotate roles in a dataset of student team conversations, and that LLMs can approximate expert annotation.
    \item We characterize role prevalence and trajectories across a semester-long project, clarifying how role dynamics change in response to specific project demands.
    \item We demonstrate that our role constructs improve downstream-task performance on peer recognition and team performance prediction across two different datasets.
\end{itemize}

\section{Related Work}

Prior studies have linked linguistic and pragmatic features (e.g., politeness, toxicity) to task success and conversational failure \citep{reitter_predicting_2007, zhang_conversations_2018}. Such interpretable features can also help predict and explain team-level performance, such as team viability \citep{cao_2021_viability}. Beyond conversational features, prior work has also used topological and structural features of communication networks \citep{Ghawi2021ImprovingTP} or static personal traits \citep[e.g., grades and MBTI,][]{personal_trait} to predict team performance.

Within this line of work, some researchers have specifically studied how roles affect communication and teamwork. \citet{banerjee-rudnicky-2006-say} predicted functional roles during meetings (e.g., \textit{leader}, \textit{scribe}) from lexical and participation cues, while \citet{non-linguistic} identified functional roles from non-linguistic cues (e.g., physical fidgeting). \citet{yang-etal-2015-weakly} and \citet{maki-etal-2017-roles} leveraged weakly supervised approaches to induce latent role representations from turn-taking behaviors and stylistic markers to predict team outcomes in MOOCs and online communities. Recent LLM-based multi-agent frameworks go further by assigning social roles to agents. Studies show that LLMs can simulate diverse character profiles \citep{lu-etal-2024-large} and exhibit emergent social behaviors, such as leadership, confrontation, and persuasion, in multi-agent environments \citep{lan-etal-2024-llm, li-etal-2025-advancing}. However, these studies approach role modeling primarily in short sessions under synthetic laboratory or simulated environments, whereas our work instead detects roles from longitudinal, real-world team conversations.

Unlike most prior role-modeling research, we ground our role definitions in organizational and behavioral science theories that conceptualize roles as communicative functions that support task and socio-emotional processes \citep{benne_functional_1948, meredith_belbin_management_2011, treo}. In engineering education, researchers have also found that appropriate role divisions benefit team-based learning from team formation to assessment \citep{work_with_who, ARANZABAL202222}. To build a role taxonomy specifically for this domain, \citet{Nestsiarovich2020TeamRA} used observation-driven studies of engineering classroom teams using interaction diagrams. Our work builds on this taxonomy by operationalizing their roles in student Slack conversations.

\section{Dataset}
\label{sec:dataset-collection}

We collect our dataset from a semester-long team project in an upper-level computer science course at a large public research university. The course topic was user interface design, and the team project accounted for 45\% of the grade. Students worked in teams of four to six, and most students were upper-year undergraduates or early graduate students.

The course was taught in person, and teams both met in person and communicated online. Each team was provided a private channel within the course Slack workspace as a default communication space for brainstorming, planning, and task updates. Teams could also coordinate using other communication channels. The project included eight graded deliverables, indexed as D1--D8: planning \& proposal (D1--D2), user research (D3), low-fidelity prototyping \& evaluation (D4--D5), and functional prototype implementation \& evaluation (D6--D8). Each deliverable window lasted one to two weeks.

At the end of the semester, 94 out of 186 enrolled students consented to share their course data for research. Our analysis focuses on consenting students whose teams actively used the course Slack channel throughout the semester (having more than 100 messages in total). We aggregate Slack messages by author and deliverable window, yielding 424 student-deliverable instances. Each instance consists of the set of messages authored by one consenting student during one deliverable window. Because consent was obtained at the individual level rather than the team level, not all members of a given team are necessarily represented in the dataset. To avoid assumptions about unobserved teammate behavior, all modeling and analyses are conducted at the individual (student-deliverable) level. This study was approved by the Institutional Review Board of the authors' university.\footnote{Raw student messages cannot be released due to privacy constraints. All data were anonymized during analysis. We share prompts and modeling details in the appendix.}

\begin{table}[t]
\centering
\small
\setlength{\tabcolsep}{6pt}
\begin{tabular}{l r}
\toprule
\textbf{Statistic} & \textbf{Value} \\
\midrule
Students analyzed / consented / enrolled & 55 / 94 / 186 \\
Teams represented & 18 \\
Student-deliverable instances & 424 \\
\midrule
Total messages & 6{,}307 \\
Total words & 93{,}474 \\
Median messages per instance & 9 \\
Median words per instance & 116 \\
\bottomrule
\end{tabular}
\caption{Dataset summary. ``Students analyzed'' refers to consenting students whose teams actively used the course Slack channel throughout the semester.}
\label{tab:dataset-stats}
\end{table}

\section{Roles in Team Communication}
\label{sec:roles}

We adopt a theory-driven role taxonomy from \citet{Nestsiarovich2020TeamRA}, which was developed via in-situ observation of engineering project teams and characterizes roles as communicative functions (e.g., initiating work, facilitating coordination, regulating participation). Compared to other popular role taxonomies designed for broader organizational settings (e.g., \citealp{meredith_belbin_management_2011, treo}), this taxonomy from education literature aligns closely with our setting of task-oriented collaboration in student project teams.

\begin{table}[t]
\centering
\small
\setlength{\tabcolsep}{6pt}
\begin{tabular}{l l}
\toprule
\textbf{Role} & \textbf{Shorthand description} \\
\midrule
Initiator & Initiate process \\
Explorer & Ask questions \\
Information Provider & Provide detailed information \\
Facilitator & Summarize / control discussion \\
Arbitrator & Solve disagreement \\
Representative & Express / answer for team \\
Gatekeeper & Fill gaps / invite others \\
Connector & Connect people/resources \\
\midrule
Passive Collector & Collect information \\
Outsider & Stay outside \\
\bottomrule
\end{tabular}
\caption{Team role taxonomy from \citet{Nestsiarovich2020TeamRA} with shorthand descriptions, complete communication patterns for each role can be found in Appendix Table~\ref{tab:roles-full}. \textit{Passive collector} and \textit{outsider} are excluded from annotation and modeling.}
\label{tab:roles-short}
\end{table}

Table~\ref{tab:roles-short} summarizes the ten roles in the original taxonomy. In our study, we focus on the eight constructive roles and exclude \textit{passive collector} and \textit{outsider}. This choice reflects the limits of the partial observation in our Slack dataset: it is difficult to infer the negative roles of low engagement from individual messages alone. Silence or low messaging can be captured more directly via explicit conversational volume statistics (e.g., message/word counts) that we model separately (more details in Section~\ref{sec:features}), but these signals do not necessarily mean that a student is an \textit{outsider} or \textit{passive collector}, since they may be heavily engaged in off-platform work (e.g., programming) or other communication channels (e.g., video meetings). We further adapt role cues to Slack-style coordination (e.g., only project-related messages count for \textit{explorer} or \textit{information provider}, rather than asking simple scheduling questions or providing availability). The full taxonomy with typical communication patterns, which serve as annotation guidelines, can be found in Appendix Table~\ref{tab:roles-full}.

\begin{table}[t]
\centering
\small
\setlength{\tabcolsep}{6pt}
\begin{tabular}{lccc}
\toprule
\textbf{Role} & \textbf{Support} & \textbf{Student Coverage}\\
\midrule
Initiator & 209 (49.3\%) & 53 (96\%)\\
Information Provider & 202 (47.6\%) & 52 (95\%)\\
Explorer & 200 (47.2\%) & 50 (91\%)\\
Facilitator & 193 (45.5\%) & 48 (87\%)\\
Gatekeeper & 92 (21.7\%) & 44 (80\%)\\
Representative & 45 (10.6\%) & 31 (56\%) \\
Connector & 32 (7.5\%) & 23 (42\%)\\
Arbitrator & 4 (0.9\%) & 4 (7\%) \\
\bottomrule
\end{tabular}
\caption{Role distribution in the expert-labeled student-deliverable instances ($N$=424), reporting how often a role is present across instances; student coverage represents how many students have enacted a role at least once throughout the semester ($N$=55).}
\label{tab:role-support}
\end{table}

\subsection{Human Role Annotation}
\label{sec:human-annotation}

\begin{figure*}[t]
  \centering
  \includegraphics[width=0.9\linewidth]{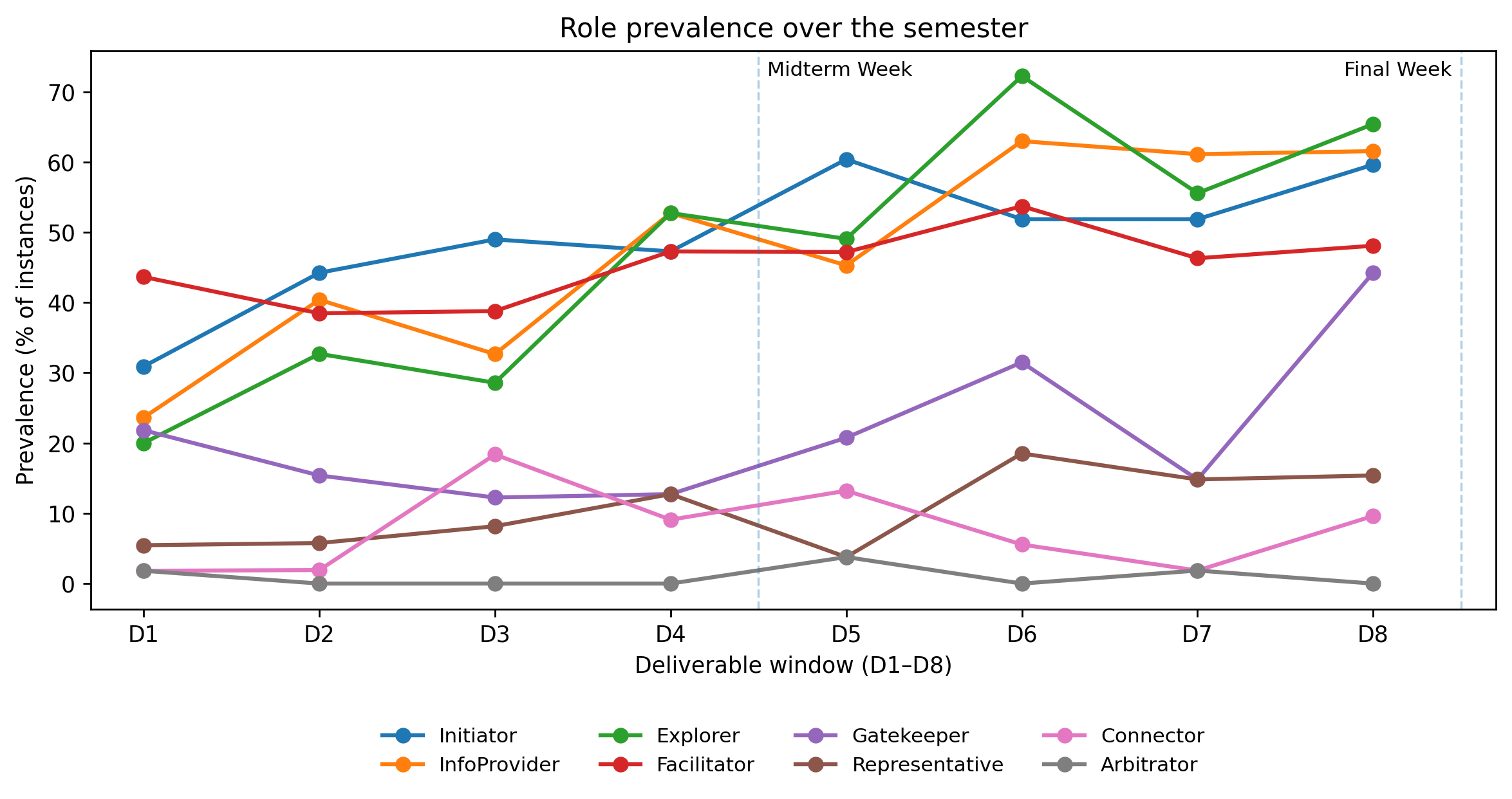}
  \caption{Role prevalence across deliverable windows. Vertical dashed lines mark midterm week (between D4 and D5) and final week (after D8). Peaks align with phase-specific demands: \textit{connector} rises during user research (D3), work/coordination roles (\textit{explorer}, \textit{information provider}, \textit{facilitator}) increase during first functional prototyping (D6), and \textit{gatekeeper} is highest during final evaluation (D8).}

  \label{fig:role-prevalence-time}
\end{figure*}

Using our adapted taxonomy, we annotate roles on each student--deliverable instance. Because roles are not mutually exclusive, we treat role labeling as a multi-label task: for each instance, annotators make an independent binary decision for each of the eight roles.

Two annotators with expertise in collaborative learning research---both are authors of this paper and former teaching staff for the course under study---labeled the dataset. We used a staged procedure to calibrate annotation and assess reliability. First, both annotators independently labeled an initial 10\% of the instances, discussed disagreements, and refined the guidelines. Next, annotators independently labeled an additional 20\% of the instances and confirmed reliability (Krippendorff’s $\alpha$ = 0.867). Disagreements in the double-labeled portion (30\% total) were adjudicated to produce a single reference label, and the remaining instances were single-annotated. 

Table~\ref{tab:role-support} summarizes the distribution of expert-annotated roles. Four work and coordination roles: \textit{initiator}, \textit{explorer}, \textit{information provider}, and \textit{facilitator} appear in roughly half of instances and most students have enacted such roles at least once, whereas \textit{gatekeeper}, \textit{representative}, and \textit{connector} are comparatively infrequent. \textit{Arbitrator} is rarely observed in these Slack logs, suggesting that explicit conflict mediation is either uncommon in this setting or more likely to occur off-platform.

\subsection{Role Dynamics}
\label{sec:role-descriptive}

To connect our role framework to concrete team behaviors, we analyze role prevalence and trajectories over deliverable windows and describe our observations below.

\paragraph{Roles shift with project phases.}
Figure~\ref{fig:role-prevalence-time} shows how role prevalence varies across the project timeline with several roles exhibiting phase-aligned peaks. \textit{Connector} spikes during the user research phase (D3), consistent with recruiting participants and reaching out to instructors/TAs or external resources when students first practice user research skills. Roles tied to technical clarification and coordination peak during the first functional prototyping window (D6): \textit{explorer}, \textit{information provider}, and \textit{facilitator} increase, aligning with the first intensive implementation and debugging cycle. \textit{Gatekeeper} rises sharply in the final evaluation window (D8), likely due to the heightened need to keep communication channels open during the busiest period of a semester---preparing finals and deadlines for all the courses. \textit{Initiator} also peaks around D5 (adjacent to midterm) and again near D8 (adjacent to final), possibly suggesting increased "driving" behaviors when time pressure is salient.

\begin{figure}[t]
  \centering
  \includegraphics[width=0.9\linewidth]{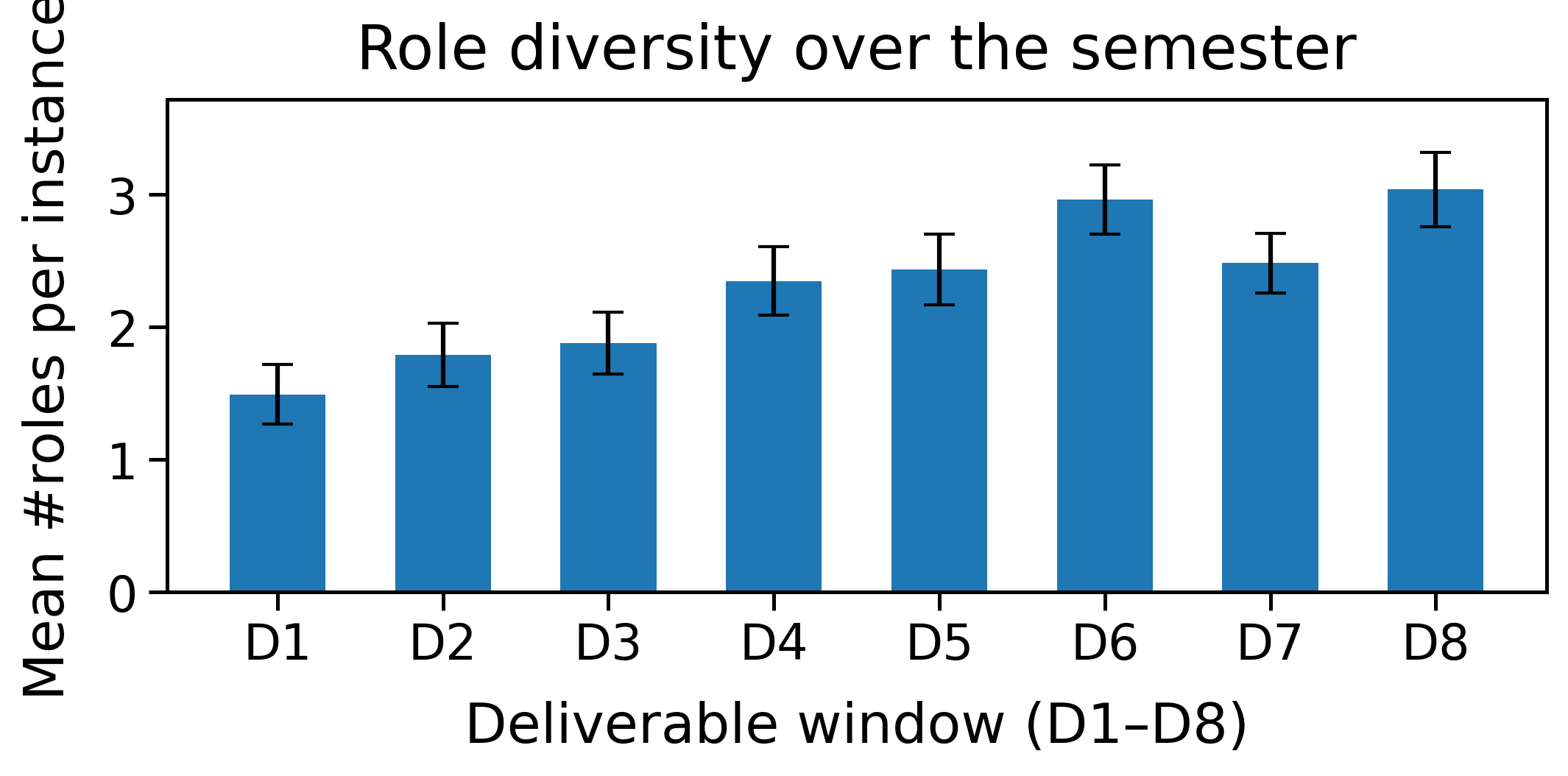}
  \caption{Mean number of roles enacted per student increases across project phases.}
  \label{fig:role-diversity-time}
\end{figure}

\paragraph{Role diversity increases over time.}
Within a single deliverable window, students typically enact multiple roles (mean = 2.3 roles per student-deliverable instance), reflecting that roles behave as situational communicative functions rather than mutually exclusive personal identities. Consistent with increasing project complexity and coordination demands, the mean number of roles enacted per student instance rises across the semester (Figure~\ref{fig:role-diversity-time}), suggesting that students begin to take on more roles as they settle into their teams. 



\subsection{LLM as Annotator}
\label{sec:llm-annotation}

\subsubsection{Methods}

Because manual role labeling does not easily scale to a large number of team interactions, we evaluate whether LLMs can replicate our expert role annotations, following recent evidence that codebook-aligned prompting can approximate human coding for complex social constructs \citep{embed_democratic_values}. We prompt the model with: 1) our adapted role definitions, 2) the ordered and anonymized messages from a student-deliverable instance, and 3) instructions to output an 8-dimensional binary role vector together with brief reasoning and evidence for each predicted positive role. We use a zero-shot prompt since few-shot generally performed worse, likely due to the nuances in chat history. The full prompt template is included in Appendix~\ref{app:role-prompt}.

\subsubsection{Results}

\begin{table}[t]
\centering
\small
\setlength{\tabcolsep}{6pt}
\begin{tabular}{lcc}
\toprule
\textbf{Role} & \textbf{F1} & \textbf{Krippendorff’s $\alpha$} \\
\midrule
Initiator & 0.916 & 0.830 \\
Explorer & 0.924 & 0.858 \\
Info Provider & 0.888 & 0.787 \\
Facilitator & 0.867 & 0.769 \\
Arbitrator & 0.750 & 0.748 \\
Representative & 0.485 & 0.414 \\
Gatekeeper & 0.712 & 0.636 \\
Connector & 0.755 & 0.738 \\
\midrule
Micro average & 0.856 & 0.799 \\
Macro average & 0.787 & 0.723 \\
\bottomrule
\end{tabular}
\caption{Best LLM performance on role labeling (GPT-5.1) against expert labels ($N$=424). We report F1 and Krippendorff’s $\alpha$.}
\label{tab:llm-annot-best}
\end{table}

We compare the model outputs against the expert labels on the same $N$ = 424 instances using per-role F1 (treating expert annotations as gold-standard labels) and Krippendorff’s $\alpha$ (treating expert annotations and LLM predictions as two coders). We experimented with multiple model families and variants, including proprietary models (GPT and Gemini) and open-source models (DeepSeek), as well as variations in model size (regular vs.\ mini) and inference mode (reasoning vs.\ non-reasoning).  Table~\ref{tab:llm-annot-best} reports results for the best-performing model variant,  GPT-5.1 (\texttt{reasoning=\{"effort":"low"\}}). The full configurations and results of cross-model comparison are reported in Appendix~\ref{app:llm-params} \& \ref{app:llm-annot-sweep}. 

When averaged across all instances, the best-performing model achieves a micro F1 of 0.856 and a micro Krippendorff’s $\alpha$ of 0.799, suggesting LLMs can be used as a scalable proxy for role annotation. When averaged at the role level, the macro averages are 0.787 and 0.723, respectively. Following the rule-of-thumb thresholds in \citet{kripp}, the four most common roles (\textit{initiator}, \textit{explorer}, \textit{information provider}, and \textit{facilitator}) yield good or excellent scores ranging from 0.77 to 0.86. The scores of less common roles (\textit{arbitrator}, \textit{gatekeeper}, and \textit{connector}), ranging from 0.64 to 0.75, are still considered acceptable, especially in LLM labeling tasks with subjective or social role labels \cite{august-etal-2020-writing, embed_democratic_values}. 

In contrast, \textit{representative} is the hardest to annotate (F1 = 0.485, Krippendorff’s $\alpha$ = 0.414), and is consistently challenging across different models, likely because "speaking for the team" often depends on conversational and social context beyond a single person’s message history. Expert annotators, as former teaching staff, may leverage additional contextual knowledge about typical teamwork processes in this course setting. While \textit{representative} only accounts for 10.6\% of the total instances, the false positives frequently co-occur with strong \textit{initiator} and \textit{facilitator} signals, suggesting the model sometimes over-interprets generic coordination or information exchange as representational behavior. This error mode is consistent with prior annotation findings that agreement varies substantially across categories and is affected by frequency \citep{wegmann-etal-2024-whats}.

\section{Task 1: Predicting Peer Recognition}
\label{sec:peer-recognition}

While role prediction itself is valuable for determining how team members communicate, we are further interested in observing the downstream benefits of roles in predicting individual and team success. Here we explore the benefits of our role constructs in two downstream tasks. We first use roles to predict peer recognition in our dataset (Section~\ref{sec:peer-recognition}), then predict team performance in an external public dataset (Section~\ref{sec:delidata}).

\subsection{Task Setup}
\label{sec:pr-task-eval}

The first task aims to predict peer recognition from teammates, represented by peer evaluation ratings collected in the same course as our Slack chat dataset. After each deliverable, students completed a peer evaluation in which each student rated each teammate on a 5-point scale (\textit{1 = below expectations, 3 = meet expectations, 5 = beyond expectations}) with optional comments. Course staff summarized received ratings into an \textit{Adjustment Factor (AF)} for each student and deliverable window. For student $i$ during deliverable window $t$,
\[
\small
\mathrm{AF}_{i,t} = 
\frac{\text{mean peer rating received by } i \text{ on } t}
{\text{mean team rating on } t}.
\]
An $\mathrm{AF}$ of $1.0$ indicates that a student is rated on par with their team’s mean for that deliverable window. Following prior work that predicts collaboration outcomes via meaningful binary splits rather than regressing a continuous score \citep{cao_2021_viability}, we frame peer recognition prediction as a binary classification task over AF splits. Based on course curriculum, course staff flagged students based on thresholds of AF:
\begin{itemize}\setlength{\itemsep}{1pt}
\small
  \item $\mathrm{AF} < 0.95$ \quad potential concerning performance;
  \item $0.95 \le \mathrm{AF} < 1.05$ \quad within the team’s typical range;
  \item $\mathrm{AF} \ge 1.05$ \quad potential outstanding performance.
\end{itemize}
Repeated occurrences with $\mathrm{AF} < 0.95$ were considered for penalty review and repeated $\mathrm{AF} \ge 1.05$ were considered for bonus review. These two thresholds, plus the average threshold of $\mathrm{AF}$ = $1.0$, naturally form meaningful splits on AF. Table~\ref{tab:class_balance} summarizes class balance. We collected the peer evaluation data from consented students. We used AF as the outcome variable because it was a real metric used in the original course, and is normalized by the team average, which controls for team-level differences in rating.

\begin{table}[t]
\centering
\small
\setlength{\tabcolsep}{6pt}
\begin{tabular}{l c c}
\toprule
\textbf{Task split} & \textbf{Positive} & \textbf{Negative} \\
\midrule
Penalty ($\mathrm{AF} < 0.95$) & 66 (15.6\%) & 358 (84.4\%) \\
Bonus ($\mathrm{AF} \ge 1.05$) & 109 (25.7\%) & 315 (74.3\%) \\
Above Avg ($\mathrm{AF} \ge 1.0$) & 218 (51.4\%) & 206 (48.6\%) \\
\bottomrule
\end{tabular}
\caption{Class distribution of three binary classification tasks for peer recognition prediction ($N$=424).}
\label{tab:class_balance}
\end{table}

\subsection{Features}
\label{sec:features}

\paragraph{Bag-of-words (BoW)}
Each instance is represented as a BoW vector over n-grams (n = 1, 2, 3) as a lexical baseline \citep{hundhausen_2023_investigating}.

\paragraph{Conversational Features}
To construct conversational baselines, we extract features for each \textit{student-deliverable instance} that are commonly used for group conversation prediction tasks from prior work.

\begin{itemize}\setlength{\itemsep}{1pt}
    \item \textbf{Volume:} Message count and word count \citep{niculae-danescu-niculescu-mizil-2016-conversational}.
    \item \textbf{Language:} Per-message polarity, subjectivity, toxicity, and readability and aggregate each score across messages using \{min, max, mean, std\} to capture a range of expression cues.\footnote{Used \texttt{TextBlob} to compute polarity and subjectivity, \texttt{Detoxify} to compute toxicity, and \texttt{textstat}'s Dale-Chall score to compute readability.} \citep{sentiment_for_stock}
    \item \textbf{Disagreement:} Count of explicit disagreement markers (e.g., "\textit{disagree}", "\textit{but}"), drawn from the ARGUE corpus \citep{allen-etal-2014-detecting}.
    \item \textbf{Interaction:} Count of direct teammate references (e.g., names and @mentions), as a proxy for social attention \citep{teamwork_dialogue}.
\end{itemize}

\paragraph{Role Features (Ours)}
We represent each instance as an 8-dimensional binary vector indicating the presence of each role. We evaluate two versions of this feature set: 1) Roles (Human) as an upper bound using expert labels; and 2) Roles (LLM) using best-performing LLM labels.

\paragraph{Role + Conversation}
To test for complementarity between high-level roles and low-level cues (e.g., a student might be an \textit{initiator} but also chat with negative sentiment), we concatenate the role vector with the conversation-feature vector.

\begin{table*}[h]
\centering
\small
\setlength{\tabcolsep}{8pt}
\begin{tabular}{lccc}
\toprule
& \multicolumn{1}{c}{\textbf{Above Avg}} & \multicolumn{1}{c}{\textbf{Penalty Risk}} & \multicolumn{1}{c}{\textbf{Bonus Potential}} \\
& \multicolumn{1}{c}{($\mathrm{AF} \ge 1.00$)} & \multicolumn{1}{c}{($\mathrm{AF} < 0.95$)} & \multicolumn{1}{c}{($\mathrm{AF} \ge 1.05$)} \\
\midrule
\textbf{Baselines} & & & \\
\quad BoW
  & 0.634 $\pm$ 0.046
  & 0.665 $\pm$ 0.034
  & 0.596 $\pm$ 0.042\\
\quad Conversational Features
  & 0.648 $\pm$ 0.029 
  & 0.640 $\pm$ 0.046 
  & 0.619 $\pm$ 0.034 \\
\midrule
\textbf{Role features (ours)} & & & \\
\quad Roles only (Human) 
  & \textbf{0.746 $\pm$ 0.015}
  & 0.741 $\pm$ 0.062
  & 0.679 $\pm$ 0.051 \\
\quad Roles only (LLM) 
  & 0.729 $\pm$ 0.027 
  & 0.730 $\pm$ 0.082 
  & 0.680 $\pm$ 0.078 \\
\quad Roles (Human) + Conversation 
  & 0.745 $\pm$ 0.012
  & \textbf{0.762 $\pm$ 0.064}
  & \textbf{0.687 $\pm$ 0.037} \\
\quad Roles (LLM) + Conversation
  & 0.723 $\pm$ 0.024
  & 0.732 $\pm$ 0.066 
  & 0.685 $\pm$ 0.058\\
\midrule
\textbf{Zero-shot LLM} & & & \\
\quad Chat only & 0.699  & 0.694 & 0.627\\
\quad Chat + Roles & 0.710  & 0.681 & 0.649\\
\bottomrule
\end{tabular}
\caption{ROC-AUC for predicting peer recognition. Supervised models report 5-fold student-grouped cross-validation mean $\pm$ std; zero-shot LLM baselines are single-pass evaluations.}
\label{tab:main-results}
\end{table*}

\subsection{Models}
\label{sec:models}

\paragraph{Supervised Classifier}
For feature sets that are already structured (BoW, conversation, role), we use Logistic Regression as the primary supervised classifier for its interpretability and stable performance on modest-sized datasets, following the approach in \citet{cao_2021_viability}. Implementation details are provided in Appendix \ref{app:peerrec-logreg}.

\paragraph{Zero-shot LLM Prompting}\setlength{\itemsep}{1pt}
Complementing the supervised classifier, we evaluate a zero-shot prompting approach to benchmark against pure LLM reasoning without explicit feature learning. We test two settings:

\begin{itemize}\setlength{\itemsep}{1pt}
    \item \textbf{Chat-only Baseline:} Raw student-deliverable message sequence as input to predict probabilities for each binary AF split. This serves as a baseline without role operationalization.
    \item \textbf{Chat + Roles:} The model is provided with both message sequence and labeled role information. This benchmarks whether the LLM's internal reasoning can leverage role information better than our supervised classifier.
\end{itemize}
Both approaches use the best-performing LLM in role annotation (GPT-5.1). Detailed prompts are reported in Appendix \ref{app:peerrec-zeroshot}.

\subsection{Evaluation}

We report ROC-AUC as the primary metric, which is robust to class imbalance, following the same setting from \citet{cao_2021_viability}. Supervised models are evaluated with 5-fold cross-validation where folds are grouped by student identity (i.e., all instances from a student appear in exactly one test fold), preventing leakage across a student’s multiple deliverable windows. We report mean $\pm$ std across folds. Zero-shot LLM baselines are evaluated once over the full dataset (no training), thus only one score is reported.

\subsection{Results}
\label{sec:pr-results}

Table~\ref{tab:main-results} reports results from our different modeling approaches. Role-based representations provide substantially stronger predictive signal than lexical and conversational baselines across all three tasks. While BoW and conversational features yield modest performance (ROC-AUC $\approx$ 0.60--0.66), role features reach 0.746 / 0.741 / 0.679 (Above Avg / Penalty / Bonus) using expert role labels, and remain competitive when roles are produced by LLM annotation (0.729 / 0.730 / 0.680). This suggests that theory-grounded roles capture structure in team communication that is not well recovered by surface-level lexical variation or generic conversation statistics alone, and that LLM-based role annotation is a viable substitute when expert coding is unavailable.

Zero-shot prompting is better than other baselines but weaker than supervised role-based models. Prompting on student chat messages with or without roles yields similar results, suggesting that role features are better used as structured constructs, rather than integrating in end-to-end prompting.

Combining roles with conversational features yields the strongest performance for penalty and bonus prediction (0.762 and 0.687, respectively). In predicting above vs. below average, role features alone perform comparably, suggesting that role indicators already capture much of the signal needed to separate above- vs.\ below-average peer recognition, while conversational statistics help more in identifying more extreme situations.

Finally, task difficulty differs by split: predicting penalty tends to be easier than predicting bonus. A plausible explanation is that under-performance cues (e.g., lack of coordination or low messaging) are more directly reflected in chat behavior, whereas exceptional contributions may also occur outside Slack (e.g., implementation work).

\section{Task 2: Predicting Team Performance}
\label{sec:delidata}

We selected a second task of predicting team-level performance using complete team dialogues, compared to the first task of predicting peer recognition at individual level. Our goal with this second task was two-fold: 1) to evaluate the generalizability of our role constructs beyond our dataset and educational context, and 2) to test our role constructs in a setting where team interaction was fully observed (i.e., all conversations were collected).


\subsection{The DeliData Corpus}
\label{sec:delidata-benchmark}

We evaluate on DeliData \citep{karadzhov2023delidata}, a public dataset of multi-party collaborative deliberation on the Wason card selection task \citep{wason_game}. DeliData contains 500 group dialogue transcripts (2--5 members per group, 1{,}579 total participants, 14{,}003 utterances). Each dialogue is paired with objective correctness before and after discussion, enabling prediction of \textit{conversational performance gain}---whether the group improves after deliberation. We frame performance gain as a binary classification task, reporting ROC-AUC under a leave-one-out cross-validation (LOOCV) setting, by following Section 7.1 of the original DeliData work.

\subsection{Role Features from LLM Annotation}
\label{sec:delidata-roles}

We apply the same role taxonomy (Section \ref{sec:roles}) and the same prompting setup using GPT-5.1 to infer participant roles from dialogue. Because DeliData provides the full dialogue transcript from all team members, we evaluate two annotation contexts:

\begin{itemize}\setlength{\itemsep}{1pt}
    \item \textbf{Individual context:} label each participant using only their own utterances (analogous to our main dataset setting).
    \item \textbf{Team context:} label each participant while providing the full team transcript.
\end{itemize}

Since the downstream label of performance gain is defined at team level, we aggregate participant-level role labels into a team-level feature vector. For a team with $n$ members, let $y_{i,r}\in\{0,1\}$ denote whether participant $i$ exhibits role $r$. We compute the proportion of role $r$ for team $t$:
\[
\small
p_{t,r}=\frac{1}{n}\sum_{i\in t} y_{i,r},
\]
This results in an $1\times8$ vector for the eight role proportions, each with a value within $[0, 1]$. This normalization makes role features comparable across different team sizes.

\subsection{Results}
\label{sec:delidata-results}

\begin{table}[t]
\centering
\small
\setlength{\tabcolsep}{6pt}
\begin{tabular}{l c}
\toprule
\textbf{Feature Set} & \textbf{ROC-AUC} \\
\midrule
\textbf{Reported by \citet{karadzhov2023delidata}} & \\
\quad (1) Deliberation Annotation & 0.53 \\
\quad (2) Interaction Features & 0.49 \\
\quad (3) Participation Dynamics & 0.61 \\
\quad (4) Conversational Statistics & 0.65 \\
\quad (1) + (2) + (3) + (4) & 0.70 \\
\midrule
\textbf{Ours} & \\
\quad (5) Zero-shot LLM Baseline & 0.68 \\
\quad (6) Roles (Individual Context) & 0.66 \\
\quad (7) Roles (Team Context) & 0.69 \\
\quad (6) + (4) & 0.72 \\
\quad (7) + (4) & \textbf{0.74} \\
\bottomrule
\end{tabular}
\caption{Predicting conversational performance gain on DeliData. (1)--(4) and their combination are reported from Table~5 of \citet{karadzhov2023delidata}. (6)--(7) use roles annotated by LLM. Rows combining roles with conversational statistics use our re-implementation.}
\label{tab:delidata}
\end{table}

Table~\ref{tab:delidata} compares our role-based features to representative baselines reported in \citet{karadzhov2023delidata}. Using role proportions alone, both variants are competitive: roles inferred from team context outperform those inferred from individual context (0.69 vs.\ 0.66), suggesting that access to the full team transcript helps the model infer roles that depend on interactional context. This confirms that our role constructs provide additional benefits when data collection covers the whole team.

Motivated by the complementarity of role and conversation features from the first task, we also re-implement the conversational-statistics feature set described in the original work (details in Appendix \ref{app:delidata-convstats}) and combine it with role features. Adding roles consistently improves performance over conversational statistics alone, and the best result comes from team-context roles + conversational statistics (0.74), exceeding the best reported combination of prior feature families (0.70). We also include a zero-shot LLM baseline that predicts performance gain directly from the group dialogue transcript (detailed prompts in Appendix~\ref{app:delidata-zeroshot}), which is competitive (0.68) but does not match the best role + conversation features model.

\section{Discussion}
\label{sec:discussion}

Our results position theory-grounded communication roles as interpretable, mid-level representations of teamwork. In this section, we discuss implications for role-based support in team-based education and, more broadly, for socially aware language technologies.

\paragraph{Role-based Chat and Learning Assistants}
Detecting roles from team chat enables role-aware feedback that encourages team members to enact missing roles, which is especially important in collaborative learning~\citep{HE2023101423}. For example, when a team contains many \textit{explorers} who question but few \textit{facilitators} who control the discussion, a role-aware chat assistant could prompt the team to summarize decisions or propose concrete next steps. Similarly, when participation appears uneven, it could encourage \textit{gatekeeper}-style check-ins (e.g., inviting input from quieter members). The same signals may also be used to assess and teach teamwork skills based on a student's long-term communication history.

\paragraph{Designing Socially Aware Agents}
Instead of relying on fixed personas or latent behaviors, agents can be guided to adopt situational communicative roles (e.g., shifting into facilitation when coordination breaks down) while preserving human ownership of the collaboration. In single human-agent interactions, team roles can help create specific types of agents for a particular task, similar to shaping their personality \cite{li-etal-2025-big5} but specifically as "teammates"; whereas in agent-agent interactions \citep[e.g., ChatDev,][]{qian-etal-2024-chatdev}, our findings that humans tend to take on multiple roles also motivate encouraging role diversity rather than homogeneous optimization (e.g., pairing initiating and facilitating behaviors). Future work can extend our study to both human-agent and agent-agent interactions, empirically testing whether assigning specific role combinations leads to higher performance in real-life or synthetic teams.

\paragraph{Multimodal Role Modeling}
Chat logs capture only one facet of collaboration. While communication is an important dimension of collaboration, it misses other crucial parts of teamwork. For example, in an engineering classroom, a student labeled as an \textit{outsider} in chat might be a "silent workhorse" contributing heavily to the codebase. Future work can combine chat history with signals in other collaboration channels and modalities when modeling team roles, such as log data from digital collaborative systems  \citep[e.g., GitHub commits and code reviews,][]{data_to_action}. Combining communication with action-based evidence could enable more accurate role inference that aligns communicative functions with technical actions.

\section{Conclusion}

In this work, we operationalize a theory-grounded taxonomy of eight communication roles to study the dynamics of teamwork. By leveraging LLMs for scalable annotation, we characterize how team members enact and shift roles across the lifecycle of a semester-long project. We demonstrate that these role representations capture critical collaborative signals, outperforming standard lexical, conversational, and zero-shot LLM baselines in predicting both peer recognition and team performance. Ultimately, our findings highlight the value of tracking communication roles to better understand, evaluate, and support effective collaboration. Future work can combine message data with additional behavioral traces (e.g., code/document edit logs) to better capture off-platform contributions and design role-aware interventions or agents for human and socially aware AI teammates.
\section*{Limitations}
\label{sec:limitations}

While we validate the role constructs on a second external benchmark (DeliData), our primary dataset comes from a single computer science course at a North American university. The relationships between roles and peer recognition in our course setting may differ across domains, cultures, and communication channels. Our dataset also includes only students who consented and teams that actively used the course-provisioned Slack channel, which may introduce selection effects, under-representing negative or sensitive interactions. 

In addition, although LLM annotation achieves high agreement with experts in our setting, results depend on prompting choices and proprietary reasoning models for best performance, and model biases related to language style may propagate into role labels (e.g., polite messages are more likely to be labeled for constructive roles). More broadly, role detection could be misused for surveillance or high-stakes evaluation. Specifically in the educational context, we intend to use it as a formative lens for team or individual reflection, rather than an automated grading tool.

\section*{Acknowledgments}
This work was supported in part by the Strategic Instructional Innovations Program (SIIP) in the Grainger College of Engineering at the University of Illinois Urbana-Champaign (UIUC). We thank our colleagues in the Language Interaction Lab and the ORCHID Lab at UIUC for their invaluable feedback. We also thank the students who shared their data for this study and the anonymous reviewers for their helpful comments.

\bibliography{latex/ref}

\twocolumn
\appendix

\begin{table*}[t]
\centering
\small
\setlength{\tabcolsep}{6pt}
\begin{tabular}{l p{0.78\textwidth}}
\toprule
\textbf{Role} & \textbf{Typical communication pattern} \\
\midrule
\textbf{Initiator} & Active participation, proposes new ideas and tasks, and introduces new directions of work. Scheduling or organizing meetings do not count. \\
\textbf{Explorer} & Active data collecting: asks general questions; requests facts, ideas, or opinions; explores alternatives; asks to clarify or specify ideas, define terms, and provide examples. Routine scheduling questions do not count.\\
\textbf{Information Provider} & Provides detailed and extensive information: takes an active part in the conversation, but mostly talks rather than listens. Providing scheduling availability does not count. \\
\textbf{Facilitator} & Defines the task or problem; suggests a method or process for accomplishing the task; provides structure; controls discussion processes; brings the group back on track. Organizing meetings may count but only when the student actively drives coordination, not merely availability updates.\\
\textbf{Arbitrator} & Encourages the group to find agreement whenever a miscommunication arises or when the group cannot come to a common position. \\
\textbf{Representative} & Verbalizes the group’s feelings, hidden problems, questions, or ideas that others are afraid to express; answers questions referred to the whole group. \\
\textbf{Gatekeeper} & Helps keep communication channels open: fills gaps in conversation; asks a person for their opinion; is sensitive to signals indicating that people want to participate. \\
\textbf{Connector} & Connects the team with people outside the group. Internal links or purely technical resources such as docs/libraries do not count. \\
\midrule
\textbf{Passive Collector} & Passive data collecting: non-verbal signs of agreement or short yes/no answers; low verbal participation in team discussion; attentive listening; keeping ideas inside (non-vocalization). \\
\textbf{Outsider} & Does not participate in project discussion.\\
\bottomrule
\end{tabular}
\caption{Full adapted team role taxonomy and role descriptions from \citet{Nestsiarovich2020TeamRA}. We retain the eight constructive roles (top block) and exclude \textit{passive collector} and \textit{outsider} for modeling in the main paper. The communication patterns are used as annotation guidelines for both human labeling and LLM prompting.}
\label{tab:roles-full}
\end{table*}

\section{Full Adapted Role Taxonomy}
We adapt the education-grounded team-role taxonomy of \citet{Nestsiarovich2020TeamRA} to Slack-based project coordination by clarifying boundary cases (e.g., routine scheduling does not count toward work roles; internal technical links do not count as \textit{connector}). Table~\ref{tab:roles-full} lists the full guidelines used for expert annotation.

\section{LLM Role Annotation Details}
\label{app:llm-prompts}

\subsection{Role annotation prompt}
\label{app:role-prompt}

We use zero-shot prompting to label the presence of each of the eight constructive roles for a student--deliverable instance (Section~\ref{sec:roles}).

\paragraph{System prompt template.}
The system message instructs the model to make role-by-role binary decisions using only explicit textual evidence in the provided messages, and to cite evidence by message indices. The following prompt skips the communication patterns of the roles, which are identical to Appendix Table~\ref{tab:roles-full}.

\vspace{0.25em}
\noindent\texttt{\small
You are an expert annotator labeling team roles using only the student's Slack messages in one deliverable period of a semester-long team project.\\
Work role-by-role. For each role, choose the final label (0 or 1) for whether the student demonstrated that role in this deliverable period and provide brief reasons (1--2 sentences) for positive labels referencing specific message numbers (e.g., [3], [10]) as evidence.\\
Do not guess from tone, use clear textual actions (e.g., asking, proposing, summarizing, mediating, representing, inviting, connecting). Multi-label allowed. Label ambiguous ones as 0. \\
Roles:\\
<< ROLE\_DEFS, see Appendix Table~\ref{tab:roles-full} >>
}

\paragraph{User prompt template.}
The user prompt includes all messages authored by one student within one deliverable window, indexed in chronological order. To reduce formatting errors, we request a structured JSON output with a fixed schema.

\vspace{0.25em}
\noindent\texttt{\small
Student messages (indexed and anonymized):\\
\{[1] xxx 
[2] xxx
...
[N] xxx\}\\
Return strict JSON matching the schema.
}

\paragraph{Output format.}
The JSON object contains exactly eight keys (one per role). Each key maps to an object with an integer \texttt{label} $\in\{0,1\}$ and a short \texttt{reason}.

\vspace{0.25em}
\noindent\texttt{\small
\{\\
\ \ "Initiator": \{"label": 0/1, "reason": "..."\},\\
\ \ "Explorer": \{"label": 0/1, "reason": "..."\},\\
\ \ \ldots\\
\ \ "Connector": \{"label": 0/1, "reason": "..."\}\\
\}
}

\subsection{Inference settings and API parameters}
\label{app:llm-params}

\begin{table*}[t]
\centering
\small
\begin{tabular}{ll}
\toprule
\textbf{Setting} & \textbf{Value} \\
\midrule
Task & Role annotation (8-way multi-label; binary per role) \\
API & OpenAI API for GPT series and OpenRouter API for Gemini and DeepSeek series\\
Input unit & One participant's messages within a deliverable window \\
Output format & JSON Schema with fields \texttt{label} and \texttt{reason} per role \\
Max output tokens & \texttt{max\_output\_tokens = 5000} \\
\midrule
\multicolumn{2}{l}{\textbf{Reasoning Setting (detailed model variants in Appendix Table \ref{tab:llm-annot-sweep})}}\\
For GPT & \texttt{Effort} can be set to \texttt{none, minimal, low, medium, high} \\
For Gemini & \texttt{Effort} can be set to \texttt{minimal, low, medium, high} \\
For DeepSeek & Can choose either \texttt{thinking.enabled} or \texttt{thinking.disabled} \\
\bottomrule
\end{tabular}
\caption{LLM role-annotation inference settings; unmentioned parameters used default values}
\label{tab:llm-params}
\end{table*}

We run role annotation via the OpenAI API for GPT models and via OpenRouter API for Gemini and DeepSeek models. Table~\ref{tab:llm-params} summarizes the inference settings used for LLM role annotation. Unless noted below, we use API default values for the parameters.

\subsection{Macro performance across LLM variants}
\label{app:llm-annot-sweep}

\begin{table*}[t]
\centering
\small
\setlength{\tabcolsep}{6pt}
\begin{tabular}{lcc}
\toprule
\textbf{Model} & \textbf{Macro F1} & \textbf{Macro Krippendorff’s $\alpha$} \\
\midrule
GPT-5-mini (\texttt{reasoning=\{"effort":"minimal"\}}) & 0.614 & 0.422 \\
GPT-5.1 (\texttt{reasoning=\{"effort":"none"\}}) & 0.706 & 0.600 \\
GPT-5.1 (\texttt{reasoning=\{"effort":"low"\}}) & \textbf{0.787 }& \textbf{0.723} \\
GPT-5.1 (\texttt{reasoning=\{"effort":"high"\}}) & 0.761 & 0.671 \\
Gemini 3 Flash (\texttt{reasoning=\{"effort":"minimal"\}}) & 0.688 & 0.496 \\
Gemini 3 Pro (\texttt{reasoning=\{"effort":"low"\}}) & 0.727 & 0.576 \\
Gemini 3 Pro (\texttt{reasoning=\{"effort":"high"\}}) & 0.743 & 0.634 \\
DeepSeek V3.1 (\texttt{thinking=\{"type":"disabled"\}}) & 0.600 & 0.455 \\
DeepSeek V3.1 (\texttt{thinking=\{"type":"enabled"\}}) & 0.647 & 0.526 \\
\bottomrule
\end{tabular}
\caption{Macro performance across LLM variants for role annotation on the expert-labeled set ($N$ = 424). GPT-5.1 (\texttt{reasoning=\{"effort":"low"\}}) yields the best performance.}
\label{tab:llm-annot-sweep}
\end{table*}

We compare LLM-generated role labels against expert annotations on the same expert-labeled set. Table~\ref{tab:llm-annot-sweep} reports macro-averaged F1 and Krippendorff's $\alpha$. Across families, larger and reasoning-capable variants tend to perform better, though increasing reasoning effort is not always beneficial.

\section{Peer Recognition Experiment Details}
\label{app:peerrec-details}

\subsection{Supervised classifier configuration}
\label{app:peerrec-logreg}

All supervised peer recognition models use \texttt{scikit-learn} Logistic Regression with default
L2 regularization.

For structured feature vectors (conversational, roles, roles+conversational), we standardize features
and fit logistic regression:

\vspace{0.25em}
\noindent\texttt{\small
make\_pipeline(\\
\ \ StandardScaler(with\_mean=True),\\
\ \ LogisticRegression(solver="lbfgs", max\_iter=3000, class\_weight="balanced", random\_state=42)\\
)
}

For BoW, we use \texttt{CountVectorizer} with \texttt{ngram\_range=(1,3)} followed by logistic regression:

\vspace{0.25em}
\noindent\texttt{\small
make\_pipeline(\\
\ \ CountVectorizer(ngram\_range=(1,3)),\\
\ \ LogisticRegression(max\_iter=2000, class\_weight="balanced")\\
)
}

\subsection{Zero-shot LLM baseline prompts}
\label{app:peerrec-zeroshot}

We evaluate a zero-shot LLM baseline that predicts whether an instance satisfies a given AF threshold.
We run one independent prompt per split: penalty, bonus, or average.

\paragraph{System prompt template.}
\vspace{0.25em}
\noindent\texttt{\small
You are an expert in team communication analytics. Given one student's Slack message history,
predict the probability of the following outcome for their within-team peer evaluation score: \{CONDITION: definition of penalty, bonus, or above average, see Section~\ref{sec:pr-task-eval}\}.\\
Respond strictly as JSON with keys "prob" (0--1).
}

\paragraph{User prompt template.}
\vspace{0.25em}
\noindent\texttt{\small
Messages:\\
\{ANONYMIZED\_RAW\_TEXT\}
}

\section{DeliData Experiment Details}
\label{app:delidata-details}

\subsection{Conversational statistics features}
\label{app:delidata-convstats}

We re-implement the conversational-statistics feature set following Appendix A.3 of the original paper \cite{karadzhov2023delidata} since the authors didn't publish their code.

For each group dialogue transcript (9 features): 
\begin{itemize}
\small
\item number of participants in the chat
\item total number of messages
\item average number of messages per player
\item average number of tokens per player
\item total unique tokens
\item average unique tokens per player
\item participants’ individual performance
\item diversity in participants’ individual solutions
\item group consensus
\end{itemize}

\subsection{Zero-shot LLM baseline prompt for performance gain}
\label{app:delidata-zeroshot}

\paragraph{System prompt.}
\vspace{0.25em}
\noindent\texttt{\small
You are a team performance analyst. Given a team discussion transcript from the Wason card task,\\
predict whether the team's final performance improved compared to their initial performance.\\
Output a JSON object with a single key "prob\_gain" between 0 and 1.
}

\paragraph{User prompt.}
\vspace{0.25em}
\noindent\texttt{\small
Transcript:\\
\{TEAM\_TRANSCRIPT\}\\
\\
}

\end{document}